# Exact Maximum Entropy Inverse Optimal Control for Modeling Human Attention Switching and Control*

Felix Schmitt[1], Hans-Joachim Bieg[1], Dietrich Manstetten[1], Michael Herman[1] and Rainer Stiefelhagen[2]

*Abstract*— Maximum Causal Entropy (MCE) Inverse Optimal Control (IOC) has become an effective tool for modeling human behavior in many control tasks. Its advantage over classic techniques for estimating human policies is the transferability of the inferred objectives: Behavior can be predicted in variations of the control task by policy computation using a relaxed optimality criterion. However, exact policy inference is often computationally intractable in control problems with imperfect state observation. In this work, we present a model class that allows modeling human control of two tasks of which only one be perfectly observed at a time requiring attention switching. We show how efficient and exact objective and policy inference via MCE can be conducted for these control problems. Both MCE-IOC and Maximum Causal Likelihood (MCL)-IOC, a variant of the original MCE approach, as well as Direct Policy Estimation (DPE) are evaluated using simulated and real behavioral data. Prediction error and generalization over changes in the control process are both considered in the evaluation. The results show a clear advantage of both IOC methods over DPE, especially in the transfer over variation of the control process. MCE and MCL performed similar when training on a large set of simulated data, but differed significantly on small sets and real data.

## I. INTRODUCTION

### A. Motivation

Modeling and prediction of sequential human behavior is important in many application areas. For example, [1] recently proposed an approach to model the steering and looking behavior of automobile drivers to predict potential safety issues.

Realistic models typically require the collection of real-world behavioral data. Often the control scenario's dynamics are already known (e.g. the vehicle model) but the policy, the mapping from the system states to human control actions, has to be inferred. Classically, Direct Policy Estimation (DPE) by means of regression is employed for this purpose e.g. as in [2]. Difficulties in the application of DPE arise for system states that are not contained in the collected data. In addition, small changes in the control scenario may result in an adaptation of human behavior [3], which may pose additional problems to DPE.

An approach to avoid the limitations of DPE is provided by the concept of optimal policies in Markov Decision Processes (MDP)s. Here, the policy results from maximizing reward, which depends on the system states and agent's control actions (Optimal Control, OC[1]). Adaptivity in behavior is resulting from a reward model, which determines the specific optimal policy. For example, a static negative reward for crashes with other cars naturally results in more cautious driving behavior in dense traffic. Inverse Optimal Control (IOC) can be used to infer the reward that is underlying the observed behavior [4], [5], [6]. IOC algorithms iterate between OC and update of the reward model until the optimal behavior equals the observed behavior.

Unfortunately, many realistic human control problems can be difficult to address within a classical MDP framework. First, increased complexity due to high-dimensional state spaces can be an issue. An important MDP class to cope with this aspect are Linear Quadratic Regulators (LQRs). This MDP variant is frequently used for efficient optimal control of technical systems (e.g. lateral vehicle dynamics). System state observability can also be an issue: For example, driving an automobile comprises of subtasks such as monitoring the road scenery and reading the speedometer. Because of the limited human field of view only the states of one task at a time can be observed and an attention[2] switching strategy is required [7], [3]. Partially Observable Markov Decision Processes (POMDPs), an extension of MDPs, address this issue by explicitly allowing for the fact that state information may be imperfect [8].

OC in POMDPs can be very difficult. In most cases computationally expensive approximation techniques must be applied. The extension of LQRs to POMDPs *without* switching of the agent's attention, Linear Quadratic Gaussian (LQG) problems, can efficiently and exactly be solved using the Kalman filter. LQGs *with* switching of attention haven been applied to human behavior [9], [10], [11]. However, OC for attention switching in LQGs[3] is considerably harder [12]. Early work solved the OC problem by exhaustive search and only recently markable progress in solution techniques has been made [13]. Sprague and Ballard [14] proposed an approach to solve a nonlinear model that includes attention switching, where each subtask is solved with full observability and a heuristic arbitration mechanism is applied afterwards.

IOC methods have been proposed for POMDPs [15], using classical techniques for OC. IOC was also applied to

*This work was part of the public project UR:BAN which was co-funded by the Federal Ministry for Economic Affairs and Energy on basis of a decision by the German Bundestag.

[1]F. Schmitt felix.schmitt@de.bosch.com, H.-J. Bieg, D. Manstetten and M. Herman are with Robert Bosch GmbH, Corporate Research, 70465 Stuttgart, Germany
[2] R. Stiefelhagen is with the Institute for Anthropomatics and Robotics, Karlsruhe Institute of Technology, Germany.

---

[1]Depended on the algorithms also known as Reinforcement Learning (RL)
[2]in the sense of *selective* attention as the prioritized allocation of *limited* sensor capabilities onto *a part* of all objects or subtasks
[3]usually referred to as measurement or sensor scheduling

model human attention switching [16]. In this approach, the technique by [14] was used to approximate a solution. However, both computational expensive OC and approximate or heuristic solution can seriously undermine the advantages of IOC over DPE in practical application: If OC is hard, this hinders the prediction of human adaptation which requires on-line policy re-computation. When using scenario-specific approximate OC in iteration, IOC can fail to estimate reward models that are transferable.

To address these issues, we present a class of LQG attention switching problems, in which policy and rewards can be inferred exactly. Our approach builds on the Maximum Causal Entropy (MCE) framework [5] and extends recent work on MCE in LQGs [17]. Although we have to restrict the model class, we can preserve the application benefits of IOC mentioned earlier. We have already successfully employed this approach for modeling attention switching and driving behavior in automobile drivers [1]. In the present article, we present our approach in more general form and also formulate the required procedures for computing of policies and reward gradients. In addition, we also provide the first performance evaluation of two IOC variants, namely MCE-IOC and MCL-IOC, which showed marked differences in performance, contrary to previous assumptions [5].

*B. Contributions*

We will first introduce our model class as a subclass of POMDPs and its reduction to an MDP in section II. In section III we review the MCE approach to IOC and its variant Maximum Causal Likelihood IOC (MCL-IOC). Algorithms for efficient policy and reward gradient computation for both MCE-IOC and MCL-IOC in the specific model class are presented in section IV. Both IOC variants and DPE are evaluated in the context of modeling driver behavior in section V. Here, we will first report results on simulated data before we revisit the empirical data, which were already presented in [1]. This allows us to investigate the prediction quality as well as the transferability to new scenarios, which are characterized by variations in the process parameters.

## II. PROBLEM FORMULATION

*A. Partially Observable Markov Decision Processes*

An MDP consists of a state $x$ in an state space $\mathcal{X}$, a control or action $u$ in a control space $\mathcal{U}$, a stochastic process model $\mathcal{P}(x_{t+1}|x_t, u_t)$ and a reward function $r(x, u)$. The objective of an *finite-horizon* MDP is to find a potentially stochastic policy/controller $\pi(u|x)$ that maximizes the expected reward $\mathbb{E}[\sum_{t=0}^{T} r(x_t, u_t)|\pi, \mathcal{P}, p_0]$ over a time horizon $T$, conditioned on the policy $\pi$, the dynamics $\mathcal{P}$ and an initial state distribution $p_0$. The optimality criterion is given by the famous Bellman equations

$$Q_T(x_T, u_T) = r_T(x_T, u_T) \quad (1)$$

$$Q_t(x_t, u_t) = r_t(x_t, u_t) + \mathbb{E}[V_{t+1}(x_{t+1})|\mathcal{P}(x_{t+1}|u_t, x_t)]$$

$$\pi_\dagger(u_t|x_t) = \arg\max_{u_t \in \mathcal{U}} Q_t(x_t, u_t) \quad (2)$$

$$V_t(x_t) = \mathbb{E}[Q_t(x_t, u_t)|\pi_\dagger(u_t|x_t)]. \quad (3)$$

Here, the function $V_t(.) : \mathcal{X} \mapsto \mathbb{R}$ is called Value-function and $Q_t(.,.) : \mathcal{X} \times \mathcal{U} \mapsto \mathbb{R}$ is called Q-function. Starting with $Q_T(x_T, u_T)$ the Bellman equations can in principle be used to optimally solve MDPs recursively. However, in practical problems $\mathcal{X}, \mathcal{U}$ are usually of large size or even continuous. This often prevents analytic computation.

POMDPs extend MDPs to states that cannot directly be observed. Instead the agent relies on observations $o_t$ that result from the "true" state by an Observation Model (OM) $p^o(o_t|x_t)$. The observations $o_t$ can be used to maintain an estimate of the state $x_t$ in form of a belief distribution $b(.)$ that encodes the agents uncertainty about the true states: Given a new observation $o_t$ the belief is updated by

$$b(x_t) \propto p^o(o_t|x_t)\mathbb{E}[\mathcal{P}(x_t|x_{t-1}, u_{t-1})|b(x_{t-1}), u_{t-1}], \quad (4)$$

using the past belief $b(x_{t-1})$ and the past applied control $u_{t-1}$. Defining states $x' = b(.)$[4] and a new reward $r'(x') = r'(b(.)) := \mathbb{E}[r(x)|b(x)]$, every POMDP can formally be transformed into an equivalent MDP in the belief states [8].

*B. Problem Class*

We consider a control scenario, where there are two distinct tasks that have concurring rewards $r^p = \theta_p^\top \varphi^p$ and $r^s = \theta_s^\top \varphi^s$ Fig. 1.

The *primary* task is characterized by a process in continuous states $x^p, u^p$ and linear affine dynamics, that require continuous control input. If the agent pays attention to the primary task, the state can perfectly be observed and controls applied accordingly. In addition to that, there is a *secondary* task in discrete states and controls $x^s, u^s$. The dynamics of the secondary task are also dependent on whether the agent pays attention to it, e.g. high reward states are only reachable in this case. If attention is dedicated to the secondary task, the amount of information obtained from the primary task is reduced by a noisy observation model resulting in a decrease in control performance. The agent's attention is modeled by an observation model state $x^o$ where switches by observation models controls $u^o$ come with costs $r^o = \theta_o \varphi^o$.

Specifically, we consider the POMDPs of maximization of

$$\mathbb{E}\Big[\sum_{t=0}^{T} \theta^\top \varphi(x_t, u_t) \Big| \mathcal{P}(x_{t+1}|x_t, u_t), p^o(o_t, |x_t), p_0\Big] \quad (5)$$

$$x_t = (x_t^p, x_t^s, x_t^o)^\top \quad u_t = (u_t^p, u_t^s, u_t^o)^\top$$

$$\theta^\top \varphi(x_t, u_t) = \theta_p^\top \varphi^p(x_t^p, u_t^p) + \theta_s^\top \varphi^s(x_t^s, u_t^s) + \theta_o \varphi^o(u_t^o),$$

were process and reward models consist of:
- A time-varying linear-affine model in $x^p$ and $u^p$

$$x_{t+1}^p = A_t x_t^p + B_t u_t^p + a_t + \epsilon^p, \quad (6)$$

with system matrices $A_t, B_t, a_t$ and normally distributed noise $\mathcal{N}(\epsilon^p|\mathbf{0}, \Sigma_n^p)$ with noise covariance $\Sigma^p$.
- A negative quadratic reward on $x^p, u^p$:

$$\theta_p^\top \varphi_p(x_t^p, u_t^p) = (x_t^p)^\top \Theta_1 x_t^p + (u_t^p)^\top \Theta_2 u_t^p, \quad (7)$$

---
[4]Note that the belief state $x'$ has a continuous state space.

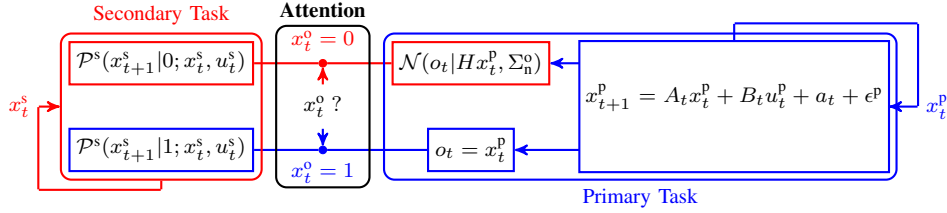

Fig. 1. Illustration of the considered dual-task problem

where $\Theta_1$ is negative semidefinite and $\Theta_2$ is negative definite.

- A linear Gaussian OM on the state $x_t^p$ which depends on the observation model state $x_t^o \in \{0,1\}$

$$p^o(o_t|x_t^p, x_t^o) = \begin{cases} \mathcal{N}(o_t|Hx_t^p, \Sigma_n^o) & \text{if } x_t^o = 0 \\ o_t = x_t^p & \text{else} \end{cases}, \quad (8)$$

with observation matrix $H$, and observation noise covariance $\Sigma_n^o$. $x_t^o$ can be switched by the OM control $u_t^o \in \{0,1\}$ according to the OM state dynamics

$$x_{t+1}^o = (x_t^o + u_t^o) \mod 2 \quad (9)$$

under the reward $\theta_o u_t^o$, $\theta_o < 0$, that implements a switching cost.

- A discrete sub-MDP of the secondary task, dependent of the observation model state $x_t^o$:

$$\mathcal{P}^s(x_{t+1}^s|x_t^o; x_t^s, u_t^s) \quad (10)$$

with an reward $\theta_s^\top \varphi^s(x_t^s, u_t^s)$.

### C. Transformation into Belief-MDP

Because of the linearity in (6) and (8) the belief $b_t$ of the partial observable state $x_t^p$ is a Gaussian and fully specified by its a-posterior expectation $\mu_t^p$ and covariance $\Sigma_t^p$. Hence, these POMDPs can be transformed into belief MDPs using the Kalman filter. Given $u_t^p, \mu_t^p, \Sigma_t^p$, (4) results in distributions

$$p(\Sigma_{t+1}^p | \Sigma_t^p) = \hat{\Sigma}_{t+1}^p - \Sigma_{t+1}^\mu \quad (11)$$
$$p(\mu_{t+1}^p|\mu_t^p, \Sigma_t^p) = \mathcal{N}(\mu_{t+1}^p|\mu_{t+1}^\mu, \Sigma_{t+1}^\mu) \quad (12)$$
$$\hat{\Sigma}_{t+1}^p := A_t \Sigma_t^p A_t^\top + \Sigma_n^p \quad (13)$$
$$K_{t+1} := \hat{\Sigma}_{t+1}^p H_{t+1}^\top (H_{t+1}\hat{\Sigma}_{t+1}^p H_{t+1}^\top + \Sigma_n^o)^{-1} \quad (14)$$
$$\Sigma_{t+1}^\mu := K_{t+1} H_{t+1} \hat{\Sigma}_{t+1}^p \quad (15)$$
$$\mu_{t+1}^\mu := A_t \mu_t^p + B u_t^p + a_t. \quad (16)$$

The expected reward under the belief $\mathbb{E}[\theta_p^\top \varphi_p(x_t^p, u_t^p)|b_t]$ is given by $(\mu_t^p)^\top \Theta_1 \mu_t^p + \mathrm{tr}(\Theta_1 \Sigma_t^p) + (u_t^p)^\top \Theta_2 u_t^p$, where tr denotes the sum of the diagonal elements of a matrix. As the equation for $K_{t+1}$ is nonlinear in $\hat{\Sigma}_{t+1}^p$ while $u^o$ is discrete, the joint optimization wrt. $\hat{\Sigma}_{t+1}^p, u_t^0$ is a nonlinear mixed-integer optimization problem. This makes exact optimal measurement scheduling in linear quadratic Gaussian problems *in general* hard [13]. In the class considered in this work, however, an efficient and exact solution is possible: As the state $x_t^p$ is fully observable if $x_t^o = 1$, the steps $d_t \in \{0, 1, \ldots, d_{\max}\}$ passed since the last exact observation can substitute the covariance $\Sigma_t^p$ as well as $x_t^o$. This enables efficient computations in IV.

### III. INVERSE OPTIMAL CONTROL

IOC seeks the reconstruction of the reward model $r(x_t, u_t)$ underlying observed optimal behavior. Here, the behavioral data $\mathcal{D}$ is given by several sequences $\{(u_t^i, x_t^i)_{t=0,\ldots,T}\}$, $i = 1, \ldots, n$ produced by an *unknown* $\pi$ under *known* initial state $p_0$ and process model $\mathcal{P}$.

#### A. Reward Reconstruction Criterion

In the case of an MDP where the reward is a linear combination of reward features $r(x_t, u_t) = \theta^\top \varphi(x_t, u_t)$, matching the empirical feature expectation

$$\underbrace{\frac{1}{n}\sum_{i=1}^n \sum_{t=0}^T \varphi(x_t^i, u_t^i)}_{=:E} \stackrel{!}{=} \mathbb{E}\Big[\sum_{t=0}^T \varphi(x_t, u_t)\Big|\pi_\dagger, \mathcal{P}, p_0\Big] \quad (17)$$

by the optimal policy $\pi_\dagger$ of reward $\theta_\dagger^\top \varphi(x_t, u_t)$ suffices that $\theta_\dagger^\top \varphi(x_t, u_t)$ is a reward for which the observed behavior performs equally as the optimal policy [4]. Therefore, (17) allows to infer the reward underlying the human policy.

#### B. The Maximum Causal Entropy Framework

In MCE-IOC [5] condition (17) is combined with maximization of the conditional entropy $-\mathbb{E}[\log \pi(u_t|x_t)|p(x_t)]$ of the policy under a distribution $p(x_t)$. The distribution $p(x_t)$ is given by evaluation of the policy $\pi$ under $\mathcal{P}$ and $p_0$. Together this forms the optimization problem:

$$\max_\pi \mathcal{H}_{\mathcal{P},p_0}(\pi) := -\mathbb{E}\Big[\sum_{t=0}^T \log\big(\pi(u_t|x_t)\big)\Big|\pi, \mathcal{P}, p_0\Big]$$

$$\text{s.t. } \mathbb{E}\Big[\sum_{t=0}^T \varphi(x_t, u_t)\Big|\pi, \mathcal{P}, p_0\Big] = E. \quad (18)$$

As a constrained optimization problem (18) can be solved via computation of a saddle-point $(\theta^\star, \pi^\star) = \arg\min_\theta \max_\pi \mathcal{L}(\theta, \pi)$ of its Lagrangian function:

$$\mathcal{L}(\pi, \theta) = \mathcal{H}_{\mathcal{P},p_0}(\pi) + \theta^\top \Big[\mathbb{E}\Big[\sum_{t=0}^T \varphi(x_t, u_t)\Big|\pi, \mathcal{P}, p_0\Big] - E\Big],$$

where $\theta$ is the Lagrangian multiplier of the feature matching constraint [5]. Setting $\nabla_\pi \mathcal{L}(\pi, \theta) \equiv 0$ allows to recursively

compute $\pi^\theta(u_t|x_t) = \arg\max_\pi \mathcal{L}(\theta, \pi)$ by

$$\tilde{Q}_t^\theta(u_T, x_T) = \theta^\top \varphi(u_T, x_T), \tag{19}$$

$$\tilde{V}_t^\theta(x_{t+1}) = \log \int \exp\left(\tilde{Q}_t^\theta(u_{t+1}, x_{t+1})\right) du_{t+1},$$

$$\tilde{Q}_t^\theta(u_t, x_t) = \theta^\top \varphi(u_t, x_t) + \mathbb{E}[\tilde{V}_t^\theta(x_{t+1})|\mathcal{P}(x_{t+1}|u_t, x_t)],$$

$$\pi^\theta(u_t|x_t) = \exp(\tilde{Q}_t^\theta(u_t, x_t) - \tilde{V}_t^\theta(x_t)). \tag{20}$$

Because of the similarity of the recursion (19) with the Bellman equations (further properties in [18] p. 76 ff.), the resulting $\pi^\theta$ can be interpreted as close-to-optimal policy for the reward $r(u_t, x_t) = \theta^\top \varphi(u_t, x_t)$.

### C. Reward-Learning

Classically, $\theta^\star$ is inferred in MCE by minimization of the Lagrangian dual of (18), $\mathcal{LD}(\theta) = \max_{\pi^\theta} \mathcal{L}(\pi^\theta, \theta)$,

$$\min_\theta \mathcal{O}^{\text{MCE}}(\theta) := \mathbb{E}\left[\tilde{V}_0^\theta(x_0)\big|p_0\right] - \frac{1}{n}\sum_{i=1}^n \sum_{t=0}^T \theta^\top \varphi(x_t^i, u_t^i)$$

by means of its gradient

$$\nabla_\theta^{\text{MCE}} = \mathbb{E}\left[\nabla \tilde{Q}_0^\theta(u_0, x_0)\big|\pi_0^\theta, p_0\right] - \frac{1}{n}\sum_{i=1}^n \sum_{t=0}^T \varphi(x_t^i, u_t^i),$$

[18] p. 187.
Alternatively, $\theta^\star$ is inferred in the MCL variant by minimization of the neg. log-likelihood of $\pi^\theta(u_t^i|x_t^i)$ (20) on data $\mathcal{D}$,

$$\min_\theta \mathcal{O}^{\text{MCL}}(\theta) := \frac{1}{n}\sum_{i=1}^n \sum_{t=0}^T \left[\tilde{V}_t^\theta(x_t) - \tilde{Q}_t^\theta(u_t, x_t)\right],$$

using its gradient

$$\nabla_\theta^{\text{MCL}} = \frac{1}{n}\sum_{i=1}^n \sum_{t=0}^T \left(\mathbb{E}[\nabla \tilde{Q}_t^\theta(u_t, x_t^i)|\pi_t^\theta] - \nabla \tilde{Q}_t^\theta(u_t^i, x_t^i)\right)$$

[19], [18] p. 189.
The gradient of $\tilde{Q}_t^\theta$ wrt. to $\theta$, $\nabla \tilde{Q}^\theta(u_t, x_t)$, can, following [19], recursively be computed by

$$\nabla \tilde{Q}_t^\theta = \varphi(x_t, u_t) + \mathbb{E}[\nabla \tilde{Q}_{t+1}^\theta|\pi_{t+1}^\theta, \mathcal{P}, u_t, x_t] \tag{21}$$

$$= \mathbb{E}\left[\sum_{j=t}^T \varphi(x_j, u_j)\bigg|\pi^\theta, \mathcal{P}, u_t, x_t\right]. \tag{22}$$

Effectively, MCE tries to find parameters, that $\pi^\theta$ *achieves* the observed rewards. MCL instead tries to find parameters, that $\pi^\theta$ *matches* the observed airs $(u, x)$ in the data.
MCL is more flexible than MCE as it allows extension to simultaneous estimation of $\mathcal{P}$ [19], but it does not necessarily estimate the same $\theta^\star$ on finite data. If $\mathcal{D}$ contains sufficiently many samples that the empirical dynamics $(x_{t+1}^i, u_t^i, x_t^i)$ equal $\mathcal{P}$ both gradients and therefore also the estimated $\theta$s coincide [18] p. 189. In practical application, however, one cannot expect that this is the case. We therefore empirically investigate the differences between MCE and MCL.

## IV. INFERENCE IN PROBLEM CLASS

Similar to [17] we apply MCE to the belief MDP derived in II-C. We will make use of the vertical and horizontal matrix concatenation, denoted by $[.;.]$ and $[.,.]$.

### A. Computation of $\tilde{Q}_t^\theta$

We use (19) for computation of $\tilde{Q}_t^\theta$. Here, $\tilde{Q}_t^\theta(u_t^p, u_t^o, u_t^s, \mu_t^p, d_t, x_t^s)$ can additively be split wrt. the discrete variables $u_t^o, u_t^s, d_t, x_t^s$ and the continuous variables $u_t^p, \mu_t^p$ This enables to conduct the recursion by application of the techniques of [17] for the LQG part and the techniques of [5] for the discrete part, resulting in the equations (23)-(25). Finally, a policy can be obtained as

$$\pi_t^\theta \propto \underbrace{\exp([\mu_t^p; u_t^p]^\top \Omega_t^1 [\mu_t^p; u_t^p] + [\mu_t^p; u_t^p]^\top \omega_t^2)}_{\mathcal{N}(u_t^p|F_t\mu^p + f_t, \Sigma_t^F)} \tag{26}$$

$$\cdot \exp(\tau_t(u_t^o, u_t^s, d_t, x_t^s)).$$

Please note, that the need of $\tilde{Q}_t^\theta$ for establishing the stochastic MCE-policy ([6] requires $Q_t^\theta$) *prevents* direct application of the efficient approximation technique of [13], as it only allows computation of the *sequence of optimal switches*.

### B. Computation of $\nabla \tilde{Q}_t^\theta$

Recursion (21) is used to obtain $\nabla \tilde{Q}_t^\theta$ by means of additive splitting similar as in $\tilde{Q}_t^\theta$. The gradients for the features $\varphi^o(u_t^o), \varphi^s(u_t^s, x_t^s)$ on the discrete variables $\nabla_{o,s}\tilde{Q}_t^\theta$ can be obtained by enumeration as in [19]. Defining $\mathfrak{F}_t := [I; F_{t+1}]$, $\mathfrak{T}_t := [[A_t, B_t]; [F_{t+1}A_t, F_{t+1}B_t]]$ and $\mathfrak{t}_t := [a_t; F_{t+1}a_t + f_{t+1}]$, the remaining gradients $\nabla_p \tilde{Q}_t^\theta$ for $\varphi(u_t^p, \mu_t^p)$ are computed according to (27), (28). Here, $\text{blk}(X, Y)$ is the block-diagonal matrix of $X, Y$, $\text{vec}(X)$ is the vertical concatenation of the columns of $X$ and $\otimes$ the Kronecker product.

## V. EVALUATION

For empirical evaluation of both MCE- and MCL-IOC as well as a classic DPE approach, we revisit the problem of driver modeling already considered in [1]. Here, we seek to model the behavior of a human who is trying to keep the lane (*primary task*) while interacting with the infotainment (*secondary task*) which requires glancing away from the road.

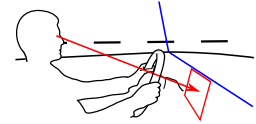

Fig. 2. Driving while interacting with infotainment.

### A. Problem Instance

We will in the following only briefly review the instance of the problem class, for a detailed derivation and explanation please refer to the original work [1].
The primary task (6) is modeled by states $x_t^p = [y_t; \dot{y}_t; \phi_t; \alpha_t]$, $u_t^p = \dot{\alpha}_t$ and dynamics given by the finite-time discretization of the kinematic vehicle model

$$\begin{bmatrix} \dot{y} \\ \dot{\phi} \\ \dot{\alpha} \end{bmatrix} = \begin{bmatrix} 0 & v_t & 0 \\ 0 & 0 & cv_t \\ 0 & 0 & 0 \end{bmatrix} \begin{bmatrix} y \\ \phi \\ \alpha \end{bmatrix} + \begin{bmatrix} 0 \\ 0 \\ 1 \end{bmatrix} \dot{\alpha} + \begin{bmatrix} 0 \\ -v_t\kappa_t \\ 0 \end{bmatrix}$$

$$\tilde{Q}_T^\theta = \underbrace{\mu_T^{\mathrm{p}\top}\Theta_1\mu_T^{\mathrm{p}} + u_T^{\mathrm{p}\top}\Theta_2 u_T^{\mathrm{p}}}_{[\mu_T^{\mathrm{p}};u_T^{\mathrm{p}}]^\top\Omega_T^1[\mu_T^{\mathrm{p}};u_T^{\mathrm{p}}]+[\mu_T^{\mathrm{p}};u_T^{\mathrm{p}}]^\top\omega_T^2+\omega_T^3} + \underbrace{\mathrm{tr}(\Theta_1\Sigma_T^{\mathrm{p}}(d_T)) + \theta_{\mathrm{o}}u_T^{\mathrm{o}} + \theta_{\mathrm{s}}^\top\varphi^{\mathrm{s}}(x_T^{\mathrm{s}},u_T^{\mathrm{s}})}_{\tau_T(u_T^{\mathrm{o}},u_T^{\mathrm{s}},d_T,x_T^{\mathrm{s}})} \quad (23)$$

$$\tilde{V}_{t+1}^\theta = \log\int\exp\left([\mu_{t+1}^{\mathrm{p}};u_{t+1}^{\mathrm{p}}]^\top\Omega_{t+1}^1[\mu_{t+1}^{\mathrm{p}};u_{t+1}^{\mathrm{p}}] + [\mu_{t+1}^{\mathrm{p}};u_{t+1}^{\mathrm{p}}]^\top\omega_{t+1}^2 + \omega_{t+1}^3\right) + \log\int\exp\left(\tau_{t+1}(u_{t+1}^{\mathrm{o}},u_{t+1}^{\mathrm{s}},d_{t+1},x_{t+1}^{\mathrm{s}})\right)$$
$$\underbrace{\phantom{}}_{\mu_{t+1}^{\mathrm{p}\top}\bar{\Omega}_{t+1}^1\mu_{t+1}^{\mathrm{p}}+\mu_{t+1}^{\mathrm{p}\top}\bar{\omega}_{t+1}^2+\bar{\omega}_{t+1}^3} \qquad \underbrace{\phantom{}}_{\bar{\tau}_{t+1}(d_{t+1},x_{t+1}^{\mathrm{s}})} \quad (24)$$

$$\bar{\Omega}_{t+1}^1 := (\Omega_{t+1}^1)_{\mu^{\mathrm{p}},\mu^{\mathrm{p}}} - (\Omega_{t+1}^1)_{\mu^{\mathrm{p}},u^{\mathrm{p}}}(\Omega_{t+1}^1)_{u^{\mathrm{p}},u^{\mathrm{p}}}^{-1}(\Omega_{t+1}^1)_{u^{\mathrm{p}},\mu^{\mathrm{p}}}, \quad \bar{\omega}_{t+1}^2 := (\omega_{t+1}^2)_{\mu^{\mathrm{p}}} - (\Omega_{t+1}^1)_{\mu^{\mathrm{p}},u^{\mathrm{p}}}(\Omega_{t+1}^1)_{u^{\mathrm{p}},u^{\mathrm{p}}}^{-1}(\omega_{t+1}^2)_{u^{\mathrm{p}}}$$

$$\bar{\omega}_{t+1}^3 := \omega_{t+1}^3 - \frac{1}{4}(\omega_{t+1}^2)_{u^{\mathrm{p}}}^\top(\Omega_{t+1}^1)_{u^{\mathrm{p}},u^{\mathrm{p}}}^{-1}(\omega_{t+1}^2)_{u^{\mathrm{p}}} + \frac{1}{2}\log(\det(-\Pi(\Omega_{t+1}^1)_{u^{\mathrm{p}},u^{\mathrm{p}}}^{-1})), \quad \Pi \text{ denotes the mathematical constant}$$

$$\tilde{Q}_t^\theta = \underbrace{\mu_t^{\mathrm{p}\top}\Theta_1\mu_t^{\mathrm{p}} + u_t^{\mathrm{p}\top}\Theta_2 u_t^{\mathrm{p}} + \mu_{t+1}^{\mu\top}\bar{\Omega}_{t+1}^1\mu_{t+1}^\mu + \mu_{t+1}^{\mu\top}\bar{\omega}_{t+1}^2 + \bar{\omega}_{t+1}^3}_{[\mu_t^{\mathrm{p}};u_t^{\mathrm{p}}]^\top\Omega_t^1[\mu_t^{\mathrm{p}};u_t^{\mathrm{p}}]+[\mu_t^{\mathrm{p}};u_t^{\mathrm{p}}]^\top\omega_t^2+\omega_t^3} \quad (25)$$
$$+ \underbrace{\mathrm{tr}(\Theta_1\Sigma_t^{\mathrm{p}}(d_t)) + \theta_{\mathrm{o}}u_t^{\mathrm{o}} + \theta_{\mathrm{s}}^\top\varphi^{\mathrm{s}}(x_t^{\mathrm{s}},u_t^{\mathrm{s}}) + \mathrm{tr}(\bar{\Omega}_{t+1}^1\Sigma_{t+1}^\mu(u_t^{\mathrm{o}},d_t)) + \mathbb{E}[\bar{\tau}_{t+1}(d_{t+1},x_{t+1}^{\mathrm{s}})|u_t^{\mathrm{o}},u_t^{\mathrm{s}},d_t,x_t^{\mathrm{s}}]}_{\tau_t(u_t^{\mathrm{o}},u_t^{\mathrm{s}},d_t,x_t^{\mathrm{s}})}$$

---

$$\nabla_p\tilde{Q}_T^\theta = \underbrace{I\mathrm{vec}([\mu_T^{\mathrm{p}};u_T^{\mathrm{p}}][\mu_T^{\mathrm{p}};u_T^{\mathrm{p}}]^\top)}_{M_T^1\mathrm{vec}([\mu_T^{\mathrm{p}};u_T^{\mathrm{p}}][\mu_T^{\mathrm{p}};u_T^{\mathrm{p}}]^\top)} + \underbrace{\mathbf{0}[\mu_T^{\mathrm{p}};u_T^{\mathrm{p}}]}_{M_T^2[\mu_T^{\mathrm{p}};u_T^{\mathrm{p}}]} + \underbrace{\mathrm{vec}(\mathrm{blk}(\Sigma_T^{\mathrm{p}}(d_T),\mathbf{0}))}_{m_T^3(u_T^{\mathrm{o}},u_T^{\mathrm{s}},d_T,x_T^{\mathrm{s}})} \quad (27)$$

$$\nabla_p\tilde{Q}_t^\theta = \underbrace{(I + M_{t+1}^1\mathfrak{T}_t\otimes\mathfrak{T}_t)\mathrm{vec}([\mu_t^{\mathrm{p}};u_t^{\mathrm{p}}][\mu_t^{\mathrm{p}};u_t^{\mathrm{p}}]^\top)}_{M_t^1\mathrm{vec}([\mu_t^{\mathrm{p}};u_t^{\mathrm{p}}][\mu_t^{\mathrm{p}};u_t^{\mathrm{p}}]^\top)} + \underbrace{\left(M_{t+1}^2\mathfrak{T}_t + M_{t+1}^1(\mathfrak{T}_t\otimes\mathbf{t}_t + \mathbf{t}_t\otimes\mathfrak{T}_t)\right)[\mu_t^{\mathrm{p}};u_t^{\mathrm{p}}]}_{M_t^2[\mu_t^{\mathrm{p}};u_t^{\mathrm{p}}]} \quad (28)$$
$$+ \underbrace{\mathrm{vec}(\mathrm{blk}(\Sigma_t^{\mathrm{p}}(d_t),\mathbf{0})) + M_{t+1}^1\mathrm{vec}(\mathbf{t}_t\mathbf{t}_t^\top + \mathfrak{F}_t\Sigma_t^\mu(u_t^{\mathrm{o}},d_t)\mathfrak{F}_t^\top + \mathrm{blk}(\mathbf{0},\Sigma_{t+1}^F)) + M_{t+1}^2\mathbf{t}_t + \mathbb{E}[m_{t+1}^3|\pi_{t+1},\mathcal{P},u_t^{\mathrm{o}},u_t^{\mathrm{s}},d_t,x_t^{\mathrm{s}}]}_{m_t^3(u_t^{\mathrm{o}},u_t^{\mathrm{s}},d_t,x_t^{\mathrm{s}})}$$

---

subject to a random disturbance $\mathcal{N}(\epsilon_{\mathrm{p}}|\mathbf{0},\Sigma_{\mathrm{n}}^{\mathrm{p}})$. In this context, $y$ is the vehicle's lateral position wrt. lane center in m, $\phi$ is the angle between tangent of the lane and the vehicle's longitudinal axis in rad, $\alpha$ is the steering-angle in rad, $v$ is the vehicle's absolute velocity in m/s, $\kappa$ is the curvature of the lane in 1/m and $c$ is the steering-wheel transmission ratio. Furthermore, primary task reward features (7) $\varphi^{\mathrm{p}}(x_t^{\mathrm{p}},u_t^{\mathrm{p}}) = [y^2;\dot{y}^2;\alpha^2;\dot{\alpha}^2]$ are used. This is combined with an observation model (8) $H = [0,0,0,1]$, $\Sigma_{\mathrm{n}}^{\mathrm{o}} = 0$. For the secondary task (10) a simple process model $x_t^{\mathrm{s}} = 1$ if $d_t > 0$, else $x_t^{\mathrm{s}} = 0$ with reward $\varphi_t^{\mathrm{s}}(x_t^{\mathrm{s}}) = x_t^{\mathrm{s}}$ is applied. Summarized, reward parameters $\theta := [\theta_1;\theta_2;\theta_3;\theta_4;\theta_5;\theta_6]$ for reward features $\varphi(x_t,u_t) := [\varphi_{1,2,3,4}^{\mathrm{p}}(x_t^{\mathrm{p}},u_t^{\mathrm{p}});\varphi^{\mathrm{s}}(u_t^{\mathrm{s}});\varphi^{\mathrm{o}}(u_t^{\mathrm{o}})]$ have to be inferred by IOC.

### B. DPE Baseline

Given the knowledge, that the policy resulting from both IOC approaches factorizes into $\pi(u_t^{\mathrm{p}}|\mu_t^{\mathrm{P}}) = \mathcal{N}(u_t^{\mathrm{p}}|F_t\mu^{\mathrm{p}} + f_t, \Sigma_t^F)$ and $\pi(u_t^{\mathrm{o}}|d_t,x_t^{\mathrm{s}})$, we use the parametrization

$$\pi_b \propto \mathcal{N}(u_t^{\mathrm{p}}|\Lambda_1^b\mu^{\mathrm{p}} + \lambda_2^b, \Sigma^b)$$
$$\cdot \exp\left(u_t^{\mathrm{o}}(\lambda_3^b d_t + \lambda_4^b x_t^{\mathrm{s}} + \lambda_5^b(1 - x_t^{\mathrm{s}}))\right), \quad (26)$$

for comparison. Time-invariant parameters were employed, as first experiments showed that this did not negatively affect performance. In every experiment, $\Lambda_1^b, \lambda_2^b, \lambda_3^b, \lambda_4^b, \lambda_5^b$ were inferred by L1- regularized maximum likelihood estimation for generalized linear models as implemented in MATLAB's `lassoglm` [20] using 5-fold cross-validation. Note that [1] uses a more domain-specific model.

### C. Numerical Experiments

In the numerical experiments a barrier $-10^{-4}\sum_{i=1}^4\log(-\theta_i^{\mathrm{p}})$ was added to the objectives $\mathcal{O}^{\mathrm{MCE,L}}(\theta)$ to ensure that $\Sigma_t^F$ is positive definite. Both optimization problems were solved to $\leq 10^{-6}$ relative gradient norm.

*1) Simulated Data:* We first conducted an evaluation $\mathcal{E}^1$ of MCE, MCL and DPE on simulated problem instances $S_1, S_2$ characterized by $v_t \equiv \{500/36, 800/36\}$ m/s, $\kappa_t = \{+14, -14\} \times 10^{-4}$ m$^{-1}$ (moderate curve on a motorway) and with initial state $x_0^{\mathrm{p}} = [0\,\mathrm{m}; 0\,\mathrm{m/s}; 0; 0]$, $d_0 = 0$, $x_0^{\mathrm{s}} = 0$. In both scenarios we first generated 3000 sequences $x^{\mathrm{p}}, x^{\mathrm{o}}, x^{\mathrm{s}}$ of $7s$ ($25\,\mathrm{Hz}$) by the MCE policies for $\theta = [-0.5\,\mathrm{m}^{-2}; -8\,\mathrm{s}^2\mathrm{m}^{-2}; -11; -200\,\mathrm{s}^2; 0.07; -3.5]$. For $k = 0, 1, \ldots, 10$, we then selected the first $2^k$ sequences of instance $S_1$, applied DPE and estimated $\theta$ using MCE and MCL. The original $\theta$ was used as initial guess for optimization in MCE and MCL. Thereafter the resulting policies were used to simulate 1976 new sequences for **both** $S_1$ and $S_2$. We report the difference between the original state distribution and the distribution obtained after policy inference. The Kullback-Leibler divergence

$$\mathrm{KL}(p(d)||p'(d)) = \sum_{d=0}^{d_{\max}} p(d)(\log[p(d)] - \log[p'(d)]),$$

was used to compare the distribution $p(d)$ in the original and obtained data. The temporal mean Kullback-Leibler divergence for Gaussians $\text{KL}^G(p(x_t^p)||p'(x_t^p))$

$$\frac{1}{2T} \sum_{t=0}^{T} \text{tr}[(\Sigma_t')^{-1}\Sigma_t'] + (\mu_t - \mu_t')^\top (\Sigma_t')^{-1}(\mu_t - \mu_t')$$
$$- \dim(x_t^p) + \log[\det(\Sigma_t')] - \log[\det(\Sigma_t)],$$

was used for primary task states. For MCE and MCL we also report the mean relative deviation of the inferred $\theta'$ from the true $\theta$, $\text{RD}(\theta, \theta') := \text{mean}_i(|\theta_i' - \theta_i|/|\theta_i|)$.

*2) Real Data:* Additionally, we evaluated on the data of [1]. They contain 1452 snippets of 5s (25 Hz) of highway driving at speeds $\{760/36, 840/36, 950/36, 1040/36\}$ m/s [5] and varying curvature $\kappa_t$. The prediction quality was assessed generating 100 sequences $x^p, x^o, x^s$ from the first states in each snippet. We first conducted 10 evaluations $\mathcal{E}^2$, where we trained on a random half of the dataset, tested on the other and conducted the procedure reversely. Thereafter we conducted 5 evaluations $\mathcal{E}^3$ where we trained on half of the data of a single speed and tested on the other half and all data of other speeds. We report the median $\text{KL}(p(d_t)||p'(d_t))$ between the true and predicted distribution $d_t$ and the median expected squared error $\text{SE}(y,y') := \mathbb{E}\left[\frac{1}{T}\sum_{t=0}^{T}(y_t - y_t')^2 | \pi, \mathcal{P}, p_0\right]$ between the true and predicted lateral position $y, y'$ over the considered snippets. $\text{SE}(y, y')$ was used instead of the $\text{KL}^G(p(x_t^p)||p'(x_t^p))$, as the KL is not defined for the single sequence per snipped.

*D. Results and Discussion*

The results of the first evaluation $\mathcal{E}^1$ on simulated data are summarized in tables I, II. In both tables we indicated the best result, i.e. the least *median* error wrt. $\text{KL}^G$, KL or RD, per condition by underlining. Additionally, we depict the results of the evaluation in Fig. 3, 4 and 5.

TABLE I
$\mathcal{E}^1$: SIMULATED DATA EVALUATION

| $\text{KL}^G$ KL | MCE $S_1$ | $S_2$ | MCL $S_1$ | $S_2$ | DPE $S_1$ | $S_2$ |
|---|---|---|---|---|---|---|
| $2^0$ | 19.89 | 19.87 | <u>19.39</u> | <u>19.54</u> | 19.90 | 606.7 |
|  | <u>0.0901</u> | <u>0.2332</u> | 0.1219 | 0.2684 | 0.1645 | 0.4057 |
| $2^4$ | 19.29 | 19.42 | <u>19.27</u> | <u>19.39</u> | 19.35 | 686.3 |
|  | 0.0041 | 0.0057 | <u>0.0018</u> | <u>0.0043</u> | 0.1653 | 0.3495 |
| $2^8$ | 19.28 | 19.40 | <u>19.22</u> | <u>19.39</u> | 19.30 | 718.2 |
|  | 0.0014 | 0.0009 | <u>0.0008</u> | <u>0.0006</u> | 0.1746 | 0.3560 |

TABLE II
$\mathcal{E}^1$: DEVIATION FROM TRUE REWARD

|  | $2^0$ | $2^2$ | $2^4$ | $2^6$ | $2^8$ | $2^{10}$ |
|---|---|---|---|---|---|---|
| MCE | 0.723 | 0.367 | <u>0.266</u> | <u>0.212</u> | 0.226 | <u>0.198</u> |
| MCL | <u>0.431</u> | <u>0.323</u> | 0.279 | 0.236 | <u>0.208</u> | 0.208 |

Similar as [4] we observed significantly[5] lower prediction error of the *best* IOC method compared to DPE on little

[5] corresponding to speeds $\{80, 90, 100, 110\}$ km/h on the speedometer

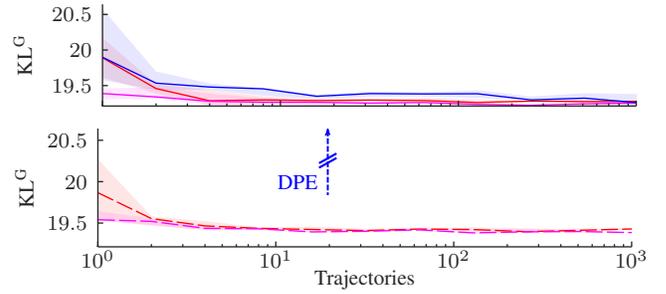

Fig. 3. $\text{KL}^G$ per number of trajectories. The medians are denoted by the line in red for MCE, in magenta for MCL and in blue for DPE. The shaded areas indicated the [0.25, 0.75] interval for MCE, MCL and DPE. The first plot with the continuous lines shows the results for $S^1$ while the second plot with the dashed lines shows the results for $S^2$.

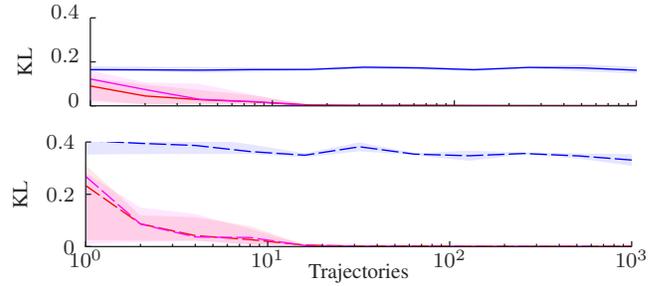

Fig. 4. KL per number of trajectories (Legend in Fig. 3).

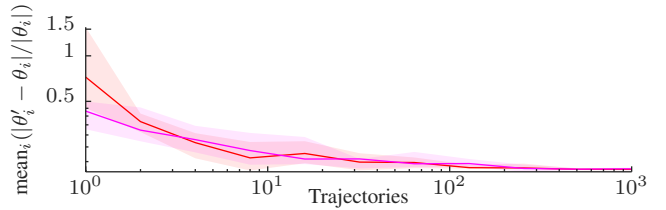

Fig. 5. RD in *log-scale* per number of trajectories (Legend in Fig. 3).

data. While DPE equaled to MCE on one trajectory, both having a higher $\text{KL}^G$ than MCL, it performed significantly worse (larger $\text{KL}^G$) from 24 to 128 trajectories. On more training data, all three methods equaled in performance and approached the average $\text{KL}^G$, 19.15, for two sets of 1976 trajectories sampled from the true policy. When evaluating of the unseen $S^2$, $\text{KL}^G$ exploded for DPE, while it only slightly increased in both IOC methods. In KL, DPE performed significantly [5] worse than both IOC methods in both $S^1$ and $S^2$. MCL and MCE show differences that, however, turned out to be statistically insignificant. Similar to $\text{KL}^G$ DPE had a significant, yet less drastically, higher KL in $S^2$. Furthermore, DPE did not reach the low KL of the IOC methods even for a large trainingset.

The results of the second evaluation $\mathcal{E}^2$ on real data are summarized in table III, while table IV presents the results of the evaluation of the transfer performance $\mathcal{E}^3$.

In both evaluations $\mathcal{E}^{2,3}$ DPE performed significantly[6] worse than both IOC methods and in $\text{KL}^G$ also worse than the domain-specific baseline of [1] (reported value of 0.045). MCE significantly[5] outperformed MCL in $\text{KL}^G$, while the differences in KL turned out insignificant. In the transfer

[6] all results were tested according to Wilcoxon signed-rank test, $p < 0.01$

TABLE III
$\mathcal{E}^2$: REAL DATA ORDINARY EVALUATION

|    | DPE Train | Test  | MCL Train | Test  | MCE Train | Test  |
|----|-----------|-------|-----------|-------|-----------|-------|
| SE | 0.095     | 0.096 | 0.021     | 0.021 | <u>0.015</u> | <u>0.015</u> |
| KL | 0.110     | 0.109 | <u>0.072</u> | <u>0.073</u> | 0.074 | 0.075 |

TABLE IV
$\mathcal{E}^3$: REAL DATA TRANSFER OVER SPEEDS

|    | DPE Same | Trans | MCL Same | Trans | MCE Same | Trans |
|----|----------|-------|----------|-------|----------|-------|
| SE | 0.088    | 0.097 | 0.021    | 0.021 | <u>0.015</u> | <u>0.015</u> |
| KL | 0.098    | 0.118 | <u>0.073</u> | <u>0.073</u> | 0.075 | 0.075 |

to other driving speeds DPE showed a significant[5] increase in both $KL^G$ while for both IOC methods no significant differences could be found.

As expected from the discussion of the properties of MCE and MCL in III-C, both methods differed on little training-data and converged when increasing the amount of data. As a single trajectory in the evaluation contained already $7\,\text{s} \times 25\,\text{Hz} = 175$ data, the better performance of MCL indicates that the log-likelihood on 175 data $(u,x)$ offers a more robust estimation criterion than the empirical feature expectation estimated from a *single* trajectory. The performance differences on real data can result from the used linear model not capturing all physical properties of the real vehicle. As a result, the observed policy might systematically deviate from the shape of the MCE/MCL policies. In that case MCE could result in lower prediction error wrt. used metrics as it focuses stronger on matching the state distribution than MCL (see III-C).

## VI. CONCLUSION

We presented a class of human attention switching and control problems with promising application potential. Although computationally intractable in similar problems, exact and efficient policy and reward inference in the MCE-IOC framework is possible in this class using the proposed recursions. Compared to DPE of the human's policy both investigated IOC methods, MCE and MCL, resulted in consistently better behavior prediction. Contrary to the expectations of previous work, however, MCL and MCE differed significantly in evaluation: MCE showed better results in the considered metrics and problem instances.

As the expressiveness of the conducted experiments is limited to the problem class, future work has to investigate the differences between MCL and MCE on other problems to understand the specific benefits of each variant. Additionally, it has to be investigated if ideas of [13] can also be applied to enable approximate IOC in attention switching, e.g. by combing it with the approximation technique of [21]. Finally, it is worth finding out if the optimization problem of MCE as presented in this work can be extended to estimation of a dynamic model as proposed for MCL [19].